\documentclass[reprint,superscriptaddress,showpacs,amsmath,amssymb,aps]{revtex4-1}
\usepackage{booktabs}
\usepackage{CJKutf8}
\usepackage{amsmath}
\usepackage{amssymb}
\usepackage{graphics}
\usepackage{graphicx}
\usepackage{dcolumn}
\usepackage[pdfpagemode=UseNone,pdfstartview=FitH,colorlinks=true,linkcolor=blue,urlcolor=blue,anchorcolor=blue,citecolor=blue]{hyperref}
\allowdisplaybreaks[4]

\newcommand{\beq}{\begin{equation}}
\newcommand{\eeq}{\end{equation}}
\newcommand{\beqa}{\begin{eqnarray}}
\newcommand{\eeqa}{\end{eqnarray}}

\def\jpc#1{{ J.\ Phys.\ C} {\bf#1}}

\def\pra#1{{ Phys.\ Rev. A\/} {\bf#1}}
\def\prb#1{{ Phys.\ Rev. B\/} {\bf#1}}

\def\prl#1{{ Phys.\ Rev.\ Lett.} {\bf#1}}

\def\pr#1{{ Phys.\ Rev.} {\bf#1}}
\def\sci#1{{ Science} {\bf#1}}

\def\nat#1{{ Nature} {\bf#1}}

\def\nac#1{{ Nature Commun.} {\bf#1}}
\def\nal#1{{ Nano Lett.} {\bf#1}}
\def\njp#1{{ New.\ J. Phys\/} {\bf#1}}

\def\sr#1{{ Sci. Rep.} {\bf#1}}

\begin{document}
\title{Spin-relaxation anisotropy in a nanowire quantum dot with strong spin-orbit coupling}
\author{Zhi-Hai\! Liu~(\begin{CJK}{UTF8}{gbsn}刘志海\end{CJK})}
\affiliation{Quantum Physics and Quantum Information Division, Beijing Computational Science Research Center, Beijing 100193, China}

\author{Rui\! Li~(\begin{CJK}{UTF8}{gbsn}李睿\end{CJK})}

\email{ruili@ysu.edu.cn}
\affiliation{Key Laboratory for Microstructural Material Physics of Hebei Province,School of Science, Yanshan University, Qinhuangdao 066004, China}

\date{\today}
\begin{abstract}
We study the impacts of the magnetic field direction on the spin-manipulation and the spin-relaxation in a one-dimensional quantum dot with strong spin-orbit coupling. The energy spectrum and the corresponding eigenfunctions in the quantum dot are obtained exactly. We find that no matter how large the spin-orbit coupling is, the electric-dipole spin transition rate as a function of the magnetic field direction always has a $\pi$ periodicity. However, the phonon-induced spin relaxation rate as a function of the magnetic field direction has a $\pi$ periodicity only in the weak spin-orbit coupling regime, and the periodicity is prolonged to $2\pi$ in the strong spin-orbit coupling regime.
\end{abstract}
\maketitle

\section{Introduction}
In recent decades, the spin-orbit couling (SOC) in III-V semiconductor materials has promoted great advances in the studies of spintronics. For instance, the pseudospin qubit in a spin-orbit coupled quantum dot is controllable by an external ac electric-field via electric-dipole spin resonant (EDSR)~\cite{Nowack2007,Fabian2008,Hu2012,Pfund2007,Golovach2006,
Petersson2012,Ruili2013,Perge2012,Perge2010,Jingtao,Rashba2003,Tokura,Khomitsky,Rashba2008}. Furthermore, a spin-orbit coupled nanowire epitaxially covered by  superconductors has been proved to be a promising system for searching the Majorana quasiparticles~\cite{Deng2016,Li2017,sau2010,Sau2012}. Thus, from the viewpoint of both the fundamental science and the practical applications, an accurate understanding of the SOC effect in quantum system becomes important.

There are two kinds of SOCs in III-V semiconductor materials: the Dresselhaus SOC generated by the bulk inversion asymmetry and the Rashba SOC induced by the structure inversion asymmetry~\cite{Winkler2003, Dresselhaus1955,Rashba1984}. Moreover, by exploiting the electric-field dependence of the Rashba SOC in semiconductor nanostructures~\cite{Liang2012,Nitta1997}, it provides a promising method for investigating the strong SOC effect in quantum system.

In semiconductor quantum dot, in order to observe the nontrivial SOC effect one should first break the time-reversal symmetry by applying an external magnetic field~\cite{Golovach2006,Sau2012}. In the presence of both the magnetic field and the SOC, only a few models are exact solvable. For example, an analytic solution for a two-dimensional (2D) quantum dot with hard-wall confining potential was given in Ref.~\cite{Sadreev2001,Gogolin2004,Intronati2013}. The exact energy spectrum and wavefunctions of a 1D square well quantum dot were given in Ref.~\cite{rui2017}. In all of the above models, the magnetic field direction is fixed. However, from both the theoretical and the experimental viewpoints, the magnetic field direction plays an important role for the observable SOC effect in quantum dot~\cite{Perge2012,Rashba2003,Ruili2013,Nowak2013,Golovach2006,Dolcini,Scarlino2014,Falko2005,Hofmann2017,Hung2017}. It is desirable to clarify the influences of the magnetic field direction on the spin properties when SOC is strong.

In this paper, we obtain exactly the eigen-energies and -functions of an electron confined in an 1D quantum dot with large SOC. Our special interest is focused on the interplay between the SOC and the magnetic field direction. When the magnetic field direction is rotated on a plane, we study both the electric-dipole transition rate and the phonon-induced relaxation rate between the lowest Zeeman sublevels. The anisotropy of the effective Land\'{e} g-factor is revealed~\cite{Tomohiro2011}. Here, in order to facilitate the study of the influence of the magnetic-field direction on the SOC effects, the original g-factor is assumed to be a constant, i.e., the corresponding bulk value. We find that no matter how large the SOC is, the Rabi frequency as a function of the magnetic field direction always has a $\pi$ periodicity. While for the phonon-induced spin relaxation rate~\cite{Borhani2012}, with the increase of the SOC, the periodicity of the relaxation rate changes from $\pi$ in the weak SOC regime to $2\pi$ in the strong SOC regime.

\section{The Model}

\begin{figure}
\centering
\includegraphics[width=0.40\textwidth]{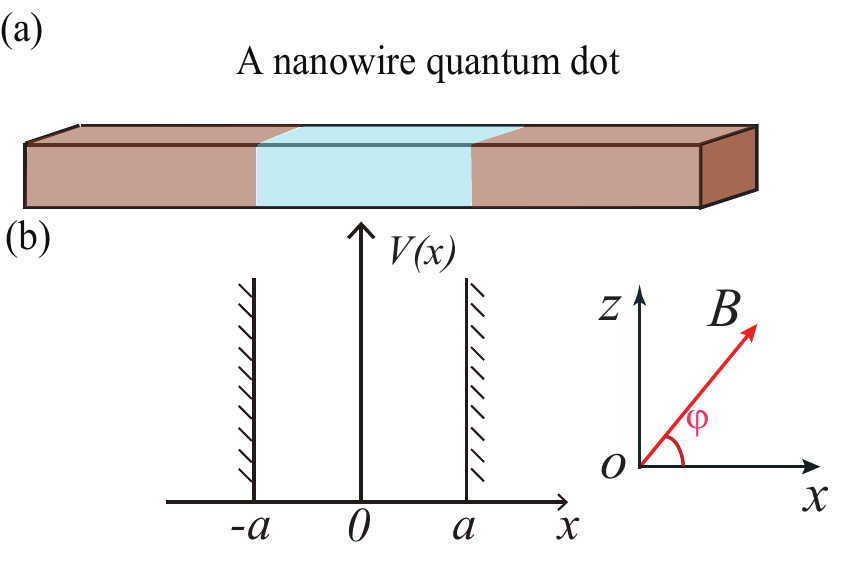}
\caption{(a) The schematic diagram of a nanowire quantum dot with large SOC. (b) The confining potential of the quantum dot along the wire axis (x-axis), and an external magnetic field $\mathbf{B}$ applied on the $x-z$ plane. }
\label{model}
\end{figure}

We consider a spin-orbit coupled quasi-1D quantum dot, where an electron confined in an infinite square well and subject to an external Zeeman field~\cite{Ruili2013}. The model under consideration is shown schematically in Fig.~\ref{model}. The nanowire  material can be chosen as those with strong SOC, e.g., InAs and InSb. Note that our approach is also applicable to materials with weak SOC.

As illustrated in Fig.~\ref{model}, the confining potential along the axial direction is modeled by an infinite square well
  \begin{align}
V(x)=\begin{cases}
 \infty, ~~~|x|>a, \\
 0,~~~~~~|x|\leq\,a,
 \end{cases}
 \label{pot}
  \end{align}
where $a$ is the half-width of the potential well. In the presence of an external magnetic field applied on the $x-z$ plane $\mathbf{B}=B(\cos\varphi,\sin\varphi)$, the Hamiltonian describing the nanowire quantum dot reads \cite{Nowak2013}
 \begin{align}
 H=\frac{p^{2}}{2m_{e}}+\alpha p\sigma^{z}+\frac{\Delta}{2}(\sigma^{x}\cos\varphi+\sigma^{z}\sin\varphi)+V(x),
 \label{hamiltonaino}
 \end{align}
where $m_{e}$ is the electron effective mass, $p=-i\hbar\partial/\partial x$ is the canonical momentum along
the wire, $\alpha$ is the Rashba SOC strength, $\Delta=g\mu_{B}B$ corresponds to the Zeeman splitting (with $g$ and $\mu_{B}$ being the Land\'{e} factor and the Bohr magneton, respectively~\cite{Wang2008,rui2017}), and $\varphi$ is the magnetic field direction. It should be noted that, in the presence of the magnetic field, there is a vector potential term  $A_{x}=-(y/2)B\sin\varphi$. However, for a quasi-1D quantum dot, we can set $y = 0$ because the motion of the electron is only allowed in the axial direction~\cite{Nowak2013,Liu2018}.

We first give the boundary condition of our model. Because the confining potential is infinite outside the well. Thus, the electron is strictly confined inside the well and the wave function is zero at the boundary sites
\begin{align}
\Psi(\pm\,a)=0,
\label{con}
\end{align}
where $\Psi(x)=[\Psi_{\uparrow}(x)~\Psi_{\downarrow}(x)]^{\rm T}$ is the eigenfunction of the quantum dot, with $\Psi_{\uparrow,\downarrow}(x)$ being its two components.

In experiments, the quantum-dot SOC depends mostly on both the material parameters and the external electric field~\cite{Winkler2003,Liang2012,Nitta1997}. As an explicit example, in our following calculations, we have chosen the InSb as our nanowire material~\cite{Sousa2003,Destefani2005}. Unless otherwise specified, the model parameters are listed in Table.~\ref{tab}.

\begin{table}
\centering
\caption{\label{tab}The relevant parameters of the InSb nanowire quantum dot we are considering. Most of the parameter values are taken from Refs.~\onlinecite{Sousa2003} and \onlinecite{Destefani2005}.}
\begin{ruledtabular}
\begin{tabular}{cccccc}
$m_{e}/m_{0}$\footnote{$m_{0}$ is the electron mass.}&$g$&$a$(nm)&$B$(T)&$x_{\rm so}$(nm)&$\varphi$ \\
$0.0136$&$-50.6$&$50$&$0.05$&40$\sim$200&0$\sim$2$\pi$  \\
 $D$(eV)&$c_{l}$(m/s)&$\rho$($\rm kg/m^{3}$)&$\rho_{l}$($\rm kg/m$)\footnote{$\rho_{l}=\rho\times \pi r_{0}^{2}$, with $r_{0}=10$nm being the radius of the nanowire.}\\
 6.6 &3690&5774.7&1.8142$\times10^{-12}$\\
\end{tabular}
\end{ruledtabular}
\end{table}

\section{the energy spectrum and the wave functions}

Inside the well, the Hamiltonian can be reduced to the following bulk Hamiltonian [$V(x)=0$ in Eq.~(\ref{hamiltonaino})]
\begin{equation}
H_{\rm b}=\frac{p^{2}}{2m_{e}}+\alpha p\sigma^{z}+\frac{\Delta}{2}(\sigma^{x}\cos\varphi+\sigma^{z}\sin\varphi).
\label{bulk}
\end{equation}
The eigenstates of the bulk Hamiltonian can be obtained by solving the bulk Schr\"{o}dinger equation $H_{\rm b}\psi(x)=E\psi(x)$.
Specifically, there are several kinds of bulk wave functions with respect to the energy region~\cite{rui2017}. In our following calculations, we focus on the energy region where only bulk plane-wave solutions are allowed.

By solving the bulk Schr\"{o}dinger equation, we find there are four independent plane-wave solutions
\begin{align}
\psi_{1,2}(x)=&e^{ik_{1,2}x}\begin{pmatrix}
\cos\phi_{1,2}  \\
\sin\phi_{1,2}
\end{pmatrix},\psi_{3,4}(x)=e^{ik_{3,4}x}\begin{pmatrix}
\sin\phi_{3,4}  \\
-\cos\phi_{3,4}
\end{pmatrix},
\label{was}
\end{align}
where $k_{1,2,3,4}$ is a function of the energy $E$ (the detailed expressions are given in Appendix~\ref{App1}) and
\begin{align}
\phi_{i}=\frac{1}{2}\arctan\left(\frac{\Delta \cos\varphi}{2\alpha\hbar k_{i}+\Delta \sin\varphi}\right).
\label{angle-phi}
\end{align}
Each independent solution does not satisfy the hard-wall boundary conditions in Eq.(\ref{con}), i.e., $\psi_{i}(\pm\,a)\neq 0$. However, a linear combination of all the degenerate bulk wave functions can fulfill the boundary condition~\cite{Sadreev2001,Gogolin2004,Intronati2013,rui2017}. Therefore,  the eigenstate of Hamiltonian (\ref{hamiltonaino}) can be written as
\begin{align}
\Psi(x)=\sum^{4}_{i=1}c_{i}\psi_{i}(x),
\label{c-m}
\end{align}
where $c_{i}$ are the coefficients to be determined. Imposing the hard-wall boundary conditions on $\Psi(x)$, we obtain an equation array for the coefficients $c_{i}$:
$\textbf{M}\cdot\textbf{C}=0$,
where $\textbf{C}=[c_{1}~c_{2}~c_{3}~c_{4}]^{\rm T}$ and the detailed expression of $\textbf{M}$ is given in Appendix~\ref{App1}. The matrix $\textbf{M}$ now is only a function of $E$. The condition that there exists nontrivial solution reads
\begin{align}
 \mathrm{Det}[\textbf{M}]=0.
 \label{det}
\end{align}
Indeed, Eq.~(\ref{det}) indicates an implicit transcendental equation for $E$, and the roots of this equation give us the energy spectrum of the quantum dot. Once the energy spectrum is obtained, we can obtain the coefficients $c_{i}$ by solving $\textbf{M}\cdot\textbf{C}=0$, such that the corresponding eigenfunctions can be obtained.

\begin{figure}
\centering
\includegraphics[width=0.38\textwidth]{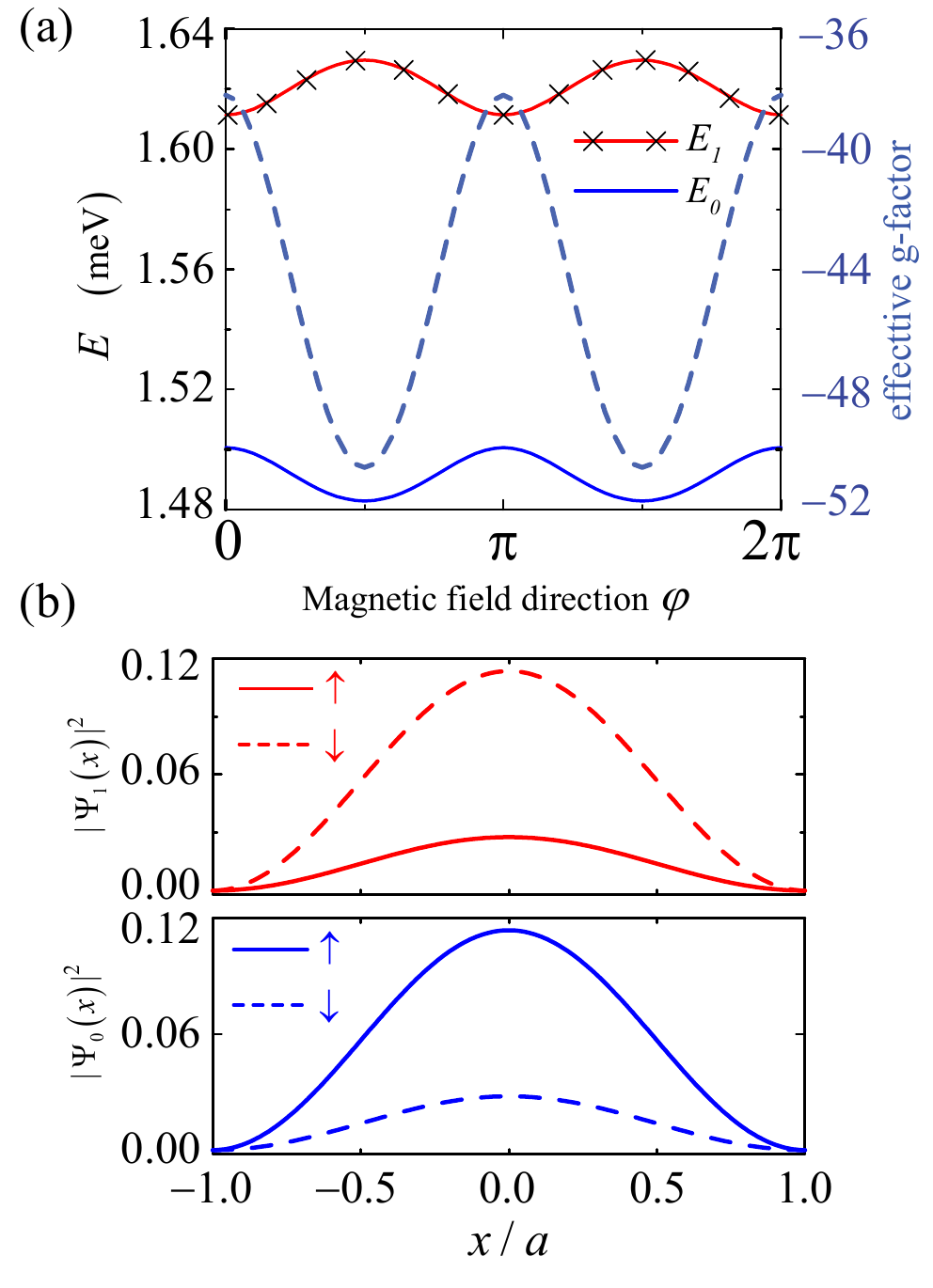}
\caption{(a) The two lowest energy levels $E_{0}$ and $E_{1}$ as a function of the magnetic field direction $\varphi$ for the SOC length $x_{\rm so}=50$ nm. The effective g-factor is defined as $g_{e}\equiv (E_{0}-E_{1})/(B\mu_{B})$. (b) The probability density distributions of the lowest two eigenstates $|\Psi_{0}(x)|^{2}$ and $|\Psi_{1}(x)|^{2}$ for magnetic direction $\varphi=\pi/6$. The solid lines represent the spin-up components and the dashed lines correspond to the spin-down components. }
\label{energy}
\end{figure}

Let $\Psi_{0}(x)$ and $\Psi_{1}(x)$ be the two lowest eigenstates in the quantum dot, and the corresponding energies are $E_{0}$ and $E_{1}$, respectively ($E_{0}<E_{1}$). When the SOC length $x_{\rm so}\equiv\hbar/(m_{e}\alpha)$ is chosen as $x_{\rm so}=50$~nm, the lowest two energy levels as a function of the magnetic field direction $\varphi$ are shown in Fig.~\ref{energy}(a). The effective g-factor $g_{e}\equiv -(E_{1}-E_{0})/(B\mu_{B})$ as a function of the angle $\varphi$ is also given. When the Zeeman field is perpendicular to the spin-orbit field, i.e., $\varphi=0$, $\pi$, and $2\pi$, the effective Zeeman splitting reaches its minimum and  $g_{e}$ becomes maximal~\cite{Nowak2013}. When the Zeeman field is parallel to the spin-orbital field, i.e., $\varphi=\pi/2$, $3\pi/2$, the effective Zeeman splitting reaches its maximum and $g_{e}$ equals to the bulk value ($g _{e}=-50.6$). We also show the probability density distribution in the quantum dot for the two lowest eigenstates $\Psi_{0}(x)$ and $\Psi_{1}(x)$ [see Fig.~\ref{energy}(b)]. As can be seen from the figure, for a general magnetic field direction $\varphi=\pi/6$, the eigenfunction contains both the spin-up component and the spin-down component. The spin-up component is dominant in the ground state and the spin-down component is dominant in the first excited state.

\section{Electric-dipole spin resonance}

In the presence of an external magnetic field, the resonant electric-dipole spin transition rate in the quantum dot was usually calculated using approximated wave functions, either the SOC or the Zeeman field was treated perturbatively~\cite{Fabian2008,Ruili2013,Golovach2006,Jingtao,Rashba2003}. Here, in our exactly solvable model, the dependence of the Rabi frequency on the magnetic filed direction is investigated.

When an alternating electric field is applied along the $x$-axis, the electric-driving Hamiltonian reads
\begin{align}
 H_{e-d}=&\frac{p^{2}}{2m_{e}}+\alpha p\sigma_{z}+\frac{g\mu_{B}}{2}\mathbf{B}\cdot\boldsymbol{\sigma}+V(x)+e\mathcal{E}x\cos\omega t,
\end{align}
where $\mathcal{E}$ and $\omega$ are the amplitude and frequency of the alternating field, respectively. Generally, under a small ac electric field the electric-dipole interaction can be regarded as a perturbation~\cite{Golovach2006,Rashba2003},  and the
resonant electric-dipole transition rate, i.e., the Rabi frequency, can be calculated:
\begin{align}
\Omega_{ij}=\frac{e\mathcal{E}}{h}\int^{a}_{-a}\Psi^{\dagger}_{i}(x) x \Psi_{j}(x)dx,
\label{omega}
\end{align}
with $h$ being the Plank constant. In the rest of this paper, we only consider the electric-dipole transition between the lowest Zeeman sublevels~\cite{Jiang2007}.

\begin{figure}
\centering
\includegraphics[width=0.36\textwidth]{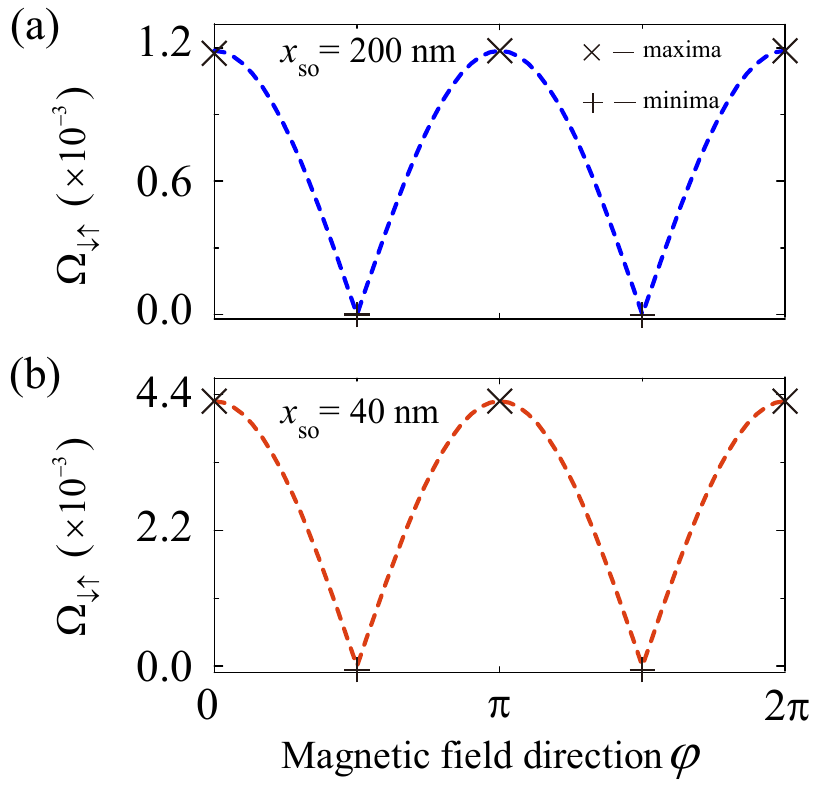}
\caption{The electric-dipole spin transition rate $\Omega_{\downarrow\uparrow}$, in unit of $e\mathcal{E}a/h$, as a function of the magnetic field direction $\varphi$, under different SOC strengths. Panel (a) show the result for SOC length $x_{\rm so}=200$~nm, and panel (b) show the result for SOC length  $x_{\rm so}=40$~nm.  }
\label{trans}
\end{figure}

The spin-flip transition rate $\Omega_{\downarrow\uparrow}$, in unit of $e\mathcal{E}a/h$, as a function of the magnetic direction is shown in Fig.~\ref{trans}. Figure~\ref{trans}(a) shows the result in the weak SOC regime $x_{\rm so}>a$, and Fig.~\ref{trans}(b) shows the result in the strong SOC regime $x_{\rm so}<a$~\cite{Liu2011}. When the Zeeman field is perpendicular to the spin-orbital field, the spin and the orbital degrees of freedom are hybridized to maximal, such that when $\varphi=0$, $\pi$, and $2\pi$, the Rabi frequency reaches its maximum. When the Zeeman field is parallel to the spin-orbit field, there is no mixing of the spin and the orbital degrees of freedom, i.e., the operator $\sigma^{z}$ is a conserved quantity, such that the Rabi frequency becomes zero at the sites $\varphi=\pi/2$ and $3\pi/2$. No matter how large the SOC is, we find that the Rabi frequency as a function of the magnetic field direction always has a $\pi$ periodicity [see Fig.~\ref{trans}].

\section{The phonon-induced Spin Relaxation}

On the one hand, the presence of SOC facilitates the manipulation of the electron spin, on the other hand, the SOC also mediates an spin-phonon interaction, which is harmful to the spin lifetime~\cite{Meunier2007,Pfund2007l,Khaetskii2000,Amasha2008,Kha2015,Bermeister2014,Woods2002,Jing2014,Wang2016}.
Here we study the dependence of the phonon-induced relaxation rate on the magnetic field direction in the quantum dot.

Due to the high excitation energy of the optical phonons, the phonon-induced spin relaxation in a semiconductor quantum dot is almost always caused by the acoustic phonons~\cite{Trif2008,Golovach2004,Segarra2015,Fabian2006,Olendski2007}. Moreover, for the energetically close two levels, i.e., the lowest Zeeman sublevels, the multi-phonon transition induced by anharmonic phonon terms can also be ignored~\cite{Inoshita1994,Dmitriev2014}. Generally, there are two kinds of acoustic electron-phonon (e-ph) interactions: the piezoelectric interaction and the deformation potential interaction~\cite{Wang2006,Cheng2004,Poudel2013,Prabhakar2013, Raith2012}.  For narrow-gap semiconductor materials with strong Rashba SOC and large g-factor, the phonon-induced relaxation is dominated by the deformation potential phonons~\cite{Romano2008,Alcalde2004}. The Hamiltonian describing the e-ph deformational interaction reads~\cite{Cleland}
\begin{align}
H_{e-ph}=\sum_{q}\left(\frac{\hbar}{2\rho_{l}\omega_{q}L}\right)^{1/2}e^{iqx}D|q|(b_{q}+b^{\dagger}_{-q}),
\label{e-ph}
\end{align}
where $\rho_{l}$ is the mass density of the nanowire, $L$ is the length of the nanowire, $D$ is the deformation potential coupling strength, $b$ ($b^{\dagger}$) denotes the phonon annihilation (creation) operator, $q$ and $\omega_{q}$ correspond to the wave vector and angular frequency of the acoustic wave. Thus, the total Hamiltonian describing the quantum-dot-phonon system reads
\begin{align}
H_{t}=H+H_{e-ph}+\sum_{q}\hbar\omega_{q}b^{\dagger}_{q}b_{q}.
\end{align}
The phonon-induced relaxation rate between the energy levels $i$ and $j$ can be calculated by using the Fermi golden rule~\cite{Golovach2004,Fabian2006}
\begin{align}
\Gamma_{ij}=\frac{D^{2}q^{2}}{2\rho_{l}\hbar\omega_{q}c_{l}}|W_{ij}(q)|^{2}\left[n(T)+1\right]\delta(\hbar\omega_{q}-\Delta_{ij}).
\label{relaxation-u}
\end{align}
Here $\Delta_{ij}=|E_{i}-E_{i}|$ is the energy difference between the relevant levels, $c_{l}$ is the wave velocity, $n(T)=\left[\exp(\Delta_{ij}/k_{B}T)-1\right]^{-1}$ is the average phonon number, and the electron-phonon matrix element $W_{ij}(q)$ is given by
\begin{align}
W_{ij}(q)=\int^{a}_{-a}\Psi^{\dagger}_{j}(x)e^{iqx}\Psi_{i}(x)dx.\label{eq_transi}
\end{align}
In the case of low temperature, $k_{B}T\ll\,\Delta_{ij}$, the average phonon number $n(T)\approx0$. Because the exact wave functions $\Psi_{i}(x)$ and $\Psi_{j}(x)$ are already obtained, the transition element $W_{ij}(q)$ can be calculated accurately, and hence the relaxation rate $\Gamma_{ij}$.

\begin{figure}
\centering
\includegraphics[width=0.36\textwidth]{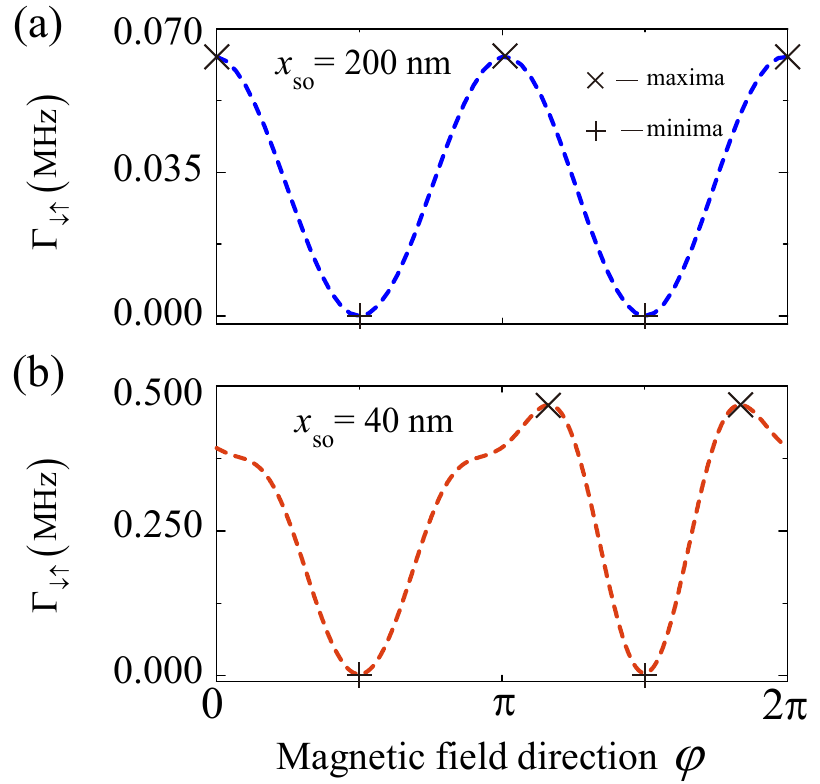}
\caption{The phonon-induced spin relaxation rate $\Gamma_{\downarrow\uparrow}$ as a function of the magnetic field direction $\varphi$, under different SOC strengths. Panel (a) shows the result in the weak SOC regime with $x_{\rm so}=200$~nm; while panel (b) shows the result in the strong SOC regime with $x_{\rm so}=40$~nm. }
\label{rela}
\end{figure}

More specifically,  the spin relaxation rate $\Gamma_{\downarrow\uparrow}$  between the two lowest energy levels as a function of the angle $\varphi$ is shown in Fig.~\ref{rela}. When the magnetic field is parallel to SOC field, i.e.,  $\varphi=\pi/2$ and $3\pi/2$, there is no spin relaxation $\Gamma_{\downarrow\uparrow}=0$ due to the fact that $\sigma^{z}$ is a good quantum number. For a relatively weak SOC $x_{\rm so}=200$ nm, at the sites $\varphi=0$, $\pi$, and $2\pi$,  the relaxation rate reaches its maximal value [see Fig.~\ref{rela}(a)]. The magnetic field dependence in this case is very similar to that of the Rabi frequency shown in Fig.~\ref{trans}. However, when the SOC is strong, i.e., $x_{\rm so}=40$ nm, we find that the sites for the maximal relaxation rate are a little bit deviation from $\varphi=\pi$ and $2\pi$ [see Fig.~\ref{rela}(b)].  This is actually a
strong SOC effect in the quantum dot. Specially, we can expand the electron-phonon operator as follows
\begin{equation}
e^{iqx}=1+iqx+\frac{1}{2}(iqx)^{2}+\cdots.
\end{equation}
When the SOC is weak, the contributions from the high-order terms ($\propto\,x^{2}$ or higher orders) to the transition element (\ref{eq_transi}) are negligible, such that the relaxation rate and the Rabi frequency share the same periodicity. When the SOC becomes strong, the contributions from the high-order terms become important, such that the sites for the maximal relaxation rate deviate from the sites of the weak SOC.  Therefore, the relaxation rate as a function of the angle $\varphi$ shows a 2$\pi$ period, in stark contrast to a $\pi$ period in the weak SOC regime~\cite{Scarlino2014,Hofmann2017,Segarra2015}.

\section{Conclusion}

In this paper, we analytically solve the 1D hard-wall quantum dot problem in the presence
of both the strong SOC and the magnetic field. The EDSR and the phonon-induced spin relaxation are studied in details with specific interest focused on the interplay between the SOC and the magnetic field direction. In different SOC regimes, we find that the phonon-induced spin relaxation rate shows different periodic oscillation over the magnetic direction. The $2\pi$ periodicity can be served as a signature of the strong SOC effect in quantum dot.

The results of our calculations will help clarify the influence of magnetic field direction on the spin-manipulation and the spin-relaxation in quantum dot under the effect of strong SOC.

\section*{Acknowledgements}

This work was supported by National Natural Science Foundation of China (grant No. 11404020) and
Postdoctoral Science Foundation of China (grant No. 2014M560039).

\begin{widetext}
\appendix

\section{The detailed expressions of $k_{i}$ and $\textbf{M}$}\label{App1}

In this Appendix, the detailed expressions of the wave vectors $k_{i}$ ($i=1,~2,~3,~4$) as a function of the energy $E$ are presented and the detailed form of the matrix $\textbf{M}$ is also given.

Expand the bulk Hamiltonian $H_{\rm b}$ in Eq.~(\ref{bulk}) in the spin space $\big\{|\uparrow\rangle,|\downarrow\rangle\big\}$,
the bulk Schr\"{o}dinger equation $H_{\rm b}\psi(x)=E\psi(x)$ can be rewritten as
\begin{align}
\begin{pmatrix}
\frac{p^{2}}{2m_{e}}+\frac{\Delta}{2}\sin\varphi+\alpha p &\frac{\Delta}{2}\cos\varphi\\
\frac{\Delta}{2}\cos\varphi&\frac{p^{2}}{2m_{e}}-\frac{\Delta}{2}\sin\varphi-\alpha p
\end{pmatrix} \psi(x)=E\psi(x),
\label{Matrix}
\end{align}
where we have used the identities: $\sigma^{z}|\uparrow\rangle=|\uparrow\rangle$, $\sigma^{z}|\downarrow\rangle=-|\downarrow\rangle$, $\sigma^{x}|\uparrow\rangle=|\downarrow\rangle$, and $\sigma^{x}|\downarrow\rangle=|\uparrow\rangle$.
 The eigenstate $\psi(x)$ is assumed to have the form of
\begin{align}
\psi(x)=e^{ikx} \begin{pmatrix}
\chi_{1}  \\
\chi_{2}
\end{pmatrix}.
\label{s-f}
\end{align}
Substituting Eq.~(\ref{s-f}) into Eq.~(\ref{Matrix}), we have
\begin{align}
\begin{pmatrix}
\frac{\hbar^{2} k^{2}}{2m_{e}}+\frac{\Delta}{2}\sin\varphi+\hbar\alpha  k -E &\frac{\Delta}{2}\cos\varphi\\
\frac{\Delta}{2}\cos\varphi&\frac{\hbar^{2} k^{2}}{2m_{e}}-\frac{\Delta}{2}\sin\varphi-\hbar\alpha k -E
\end{pmatrix}e^{ikx} \begin{pmatrix}
\chi_{1}  \\
\chi_{2}
\end{pmatrix}=0.
\label{omat}
\end{align}
Mathematically, the condition that there exists a nontrivial solution to Eq.~(\ref{omat}) reads
\begin{align}
 \mathrm{Det}\begin{pmatrix}\frac{\hbar^{2} k^{2}}{2m_{e}}+\frac{\Delta}{2}\sin\varphi+\hbar\alpha  k -E &\frac{\Delta}{2}\cos\varphi\\
\frac{\Delta}{2}\cos\varphi&\frac{\hbar^{2} k^{2}}{2m_{e}}-\frac{\Delta}{2}\sin\varphi-\hbar\alpha k -E
\end{pmatrix}=0.
 \label{K}
\end{align}
Essentially, Eq.~(\ref{K}) implies a quartic equation of $k$
\begin{align}
\frac{\hbar^{4}k^{4}}{4m_{e}^{2}}-\left(\frac{\hbar^{2}E}{m_{e}}+\alpha^{2}\hbar^{2}\right)k^{2}-\alpha\hbar\Delta\sin\varphi k+ E^{2}-\frac{\Delta^{2}}{4}=0.
\label{qua}
\end{align}
After some tedious algebra, Eq.~(\ref{qua}) can be rewritten as a product of two factor
\begin{align}
(k^{2}+k_{0} k+\eta)(k^{2}-k_{0} k+\zeta)=0,
\label{fac}
\end{align}
where
\begin{align}
\eta=(k^{3}_{0}+fk_{0}-j)/(2k_{0}),\nonumber \\
 \zeta=(k^{3}_{0}+fk_{0}+j)/(2k_{0}).
\end{align}
Here $k_{0}$ is a root of the following equation
\begin{align}
k^{6}_{0}+2fk^{4}_{0}+(f^{2}-4r)k^{2}_{0}-j^{2}=0,
\label{ko}
\end{align}
where the parameters
\begin{align}
 j=&-4\alpha m^{2}_{e}\Delta\sin\varphi/\hbar^{3}, \nonumber\\
  r=&\left(4E^{2}-\Delta^{2}\right)m^{2}_{e}/\hbar^{4}, \nonumber\\
f=&-4\left(E m_{e}+\alpha^{2}m^{2}_{e}\right)/\hbar^{2}.
\label{para}
\end{align}
One solution of Eq.~(\ref{ko}) can be written as
\begin{align}
k_{0}=\sqrt{\frac{-2f-2\sqrt{f^{2}+12r}\cos\gamma}{3}},
\label{sj}
\end{align}
where the angle
\begin{align}
\gamma=\frac{1}{3}\arccos\left[\frac{-2f^{3}+72fr-27j^{2}}{2\sqrt{(f^{2}+12r)^{3}}}\right].
\end{align}
Then, we can obtain four independent solutions to Eq.~(\ref{fac})
\begin{align}
k_{1}=&\frac{-k_{0}+\sqrt{k^{2}_{0}-4\eta}}{2},~~~k_{2}=\frac{-k_{0}-\sqrt{k^{2}_{0}-4\eta}}{2}, \nonumber\\
k_{3}=&\frac{k_{0}+\sqrt{k^{2}_{0}-4\zeta}}{2},~~~~~~~~~k_{4}=\frac{k_{0}-\sqrt{k^{2}_{0}-4\zeta}}{2},
\label{waves}
\end{align}
where the complicated dependences of the wave vectors $k_{i}$ on $E$ can be reflected by Eq.~(\ref{para}).

In the following, the detailed expression for the matrix $\textbf{M}$ is given. Substituting the eigenfunction [given in Eq.~(\ref{c-m})] into the boundary condition [see Eq.~(\ref{con})], we obtain
\begin{align}
&c_{1}e^{ik_{1}a}\cos\phi_{1}+c_{2}e^{ik_{2}a}\cos\phi_{2}
+c_{3}e^{ik_{3}a}\sin\phi_{3}+c_{4}e^{ik_{4}a}\sin\phi_{4}=0, \nonumber \\
&c_{1}e^{ik_{1}a}\sin\phi_{1}+c_{2}e^{ik_{2}a}\sin\phi_{2}
-c_{3}e^{ik_{3}a}\cos\phi_{3}-c_{4}e^{ik_{4}a}\cos\phi_{4}=0,  \nonumber \\
&c_{1} e^{-ik_{1}a}\cos\phi_{1}+c_{2} e^{-ik_{2}a}\cos\phi_{2}+c_{3} e^{-ik_{3}a}\sin\phi_{3}+c_{4} e^{-ik_{4}a}\sin\phi_{4}=0, \nonumber \\
&c_{1}e^{-ik_{1}a}\sin\phi_{1}+c_{2}e^{-ik_{2}a}\sin\phi_{2} -c_{3}e^{-ik_{3}a}\cos\phi_{3}-c_{4}e^{-ik_{4}a}\cos\phi_{4}=0.
\label{eqs}
\end{align}
The above equation array can be written as matrix equation $\textbf{M}\cdot\textbf{C}=0$, where the matrix $\textbf{M}$ reads
\begin{align}
\textbf{M}=\begin{pmatrix}
e^{ik_{1}a}\cos\phi_{1}&e^{ik_{2}a}\cos\phi_{2}
&e^{ik_{3}a}\sin\phi_{3}&e^{ik_{4}a}\sin\phi_{4}  \\
e^{ik_{1}a}\sin\phi_{1}&e^{ik_{2}a}\sin\phi_{2}
&-e^{ik_{3}a}\cos\phi_{3}&-e^{ik_{4}a}\cos\phi_{4} \\
e^{-ik_{1}a}\cos\phi_{1}&e^{-ik_{2}a}\cos\phi_{2}
&e^{-ik_{3}a}\sin\phi_{3}&e^{-ik_{4}a}\sin\phi_{4}  \\
e^{-ik_{1}a}\sin\phi_{1}&e^{-ik_{2}a}\sin\phi_{2}
&-e^{-ik_{3}a}\cos\phi_{3}&-e^{-ik_{4}a}\cos\phi_{4}
\end{pmatrix}.
\end{align}
It should be noted that $\phi_{i}$ also depends on $k_{i}$ ($i=1-4$) [see Eq.~(\ref{angle-phi})], such that matrix $\textbf{M}$ only depends on the energy $E$ [see Eqs.~(\ref{para}) and (\ref{waves})]. The condition there exists nontrivial solution for the coefficients $\textbf{C}$ reads
\begin{align}
\mathrm{Det}[\textbf{M}]=0.
\end{align}
Solving this equation, we can obtain the exact energy spectrum of the quantum dot.

\end{widetext}

\end{document}